\documentclass[aps,prl,twocolumn,superscriptaddress,showpacs]{revtex4}

\usepackage{graphicx}% Include figure files
\usepackage{dcolumn}% Align table columns on decimal point
\usepackage{bm}% bold math
\usepackage{epsfig}

\begin{document}

%===================> ADD here your LATEX definitions

\newcommand{\ee}{e$^+$e$^-$}
\newcommand{\ff}{f$_{2}$(1525)}
\newcommand{\bb}{$b \overline{b}$}
\newcommand{\cc}{$c \overline{c}$}
\newcommand{\sbs}{$s \overline{s}$}
\newcommand{\uu}{$u \overline{u}$}
\newcommand{\dd}{$d \overline{d}$}
\newcommand{\qq}{$q \overline{q}$}
\newcommand{\suo}{\rm{\mbox{$\epsilon_{b}$}}}
\newcommand{\loro}{\rm{\mbox{$\epsilon_{c}$}}}
\newcommand{\kos}{\ifmmode \mathrm{K^{0}_{S}} \else K$^{0}_{\mathrm S} $ \fi}
\newcommand{\kol}{\ifmmode \mathrm{K^{0}_{L}} \else K$^{0}_{\mathrm L} $ \fi}
\newcommand{\ko}{\ifmmode {\mathrm K^{0}} \else K$^{0} $ \fi}

\def\tpc{three-particle correlation}
\def\twopc{two-particle correlation}
\def\ksks{K$^0_S$K$^0_S$}
\def\ee{e$^+$e$^-$}
\def\ff{f$_{2}$(1525)}

\title{Two--Photon Width of $\chi_{c2}$}

\author{S.~Dobbs}
\author{Z.~Metreveli}
\author{K.~K.~Seth}
\author{A.~Tomaradze}
\author{P.~Zweber}
\affiliation{Northwestern University, Evanston, Illinois 60208}
\author{J.~Ernst}
\affiliation{State University of New York at Albany, Albany, New York 12222}
\author{K.~Arms}
\affiliation{Ohio State University, Columbus, Ohio 43210}
\author{H.~Severini}
\affiliation{University of Oklahoma, Norman, Oklahoma 73019}
\author{S.~A.~Dytman}
\author{W.~Love}
\author{S.~Mehrabyan}
\author{J.~A.~Mueller}
\author{V.~Savinov}
\affiliation{University of Pittsburgh, Pittsburgh, Pennsylvania 15260}
% ...
%\author{(CLEO Collaboration)} %FOR PRD_SPECIAL_CHANGEME
\collaboration{CLEO Collaboration} %FOR PRL,CLNS
\noaffiliation

%-------- END INSERT ------------

\begin{abstract} 
The two--photon width of $\chi_{c2}~ (^{3}P_{2})$ state of charmonium has been measured using 14.4 fb$^{-1}$ of $e^{+}e^{-}$ data taken at $\sqrt {s} = 9.46 - 11.30$ GeV with the CLEO III detector. The $\gamma\gamma$--fusion reaction studied is $e^{+}e^{-} \to e^{+}e^{-} \gamma \gamma$, 
$\gamma \gamma \to \chi_{c2} \to \gamma J/\psi \to \gamma e^+e^-(\mu^+\mu^-)$. 
We measure $\Gamma_{\gamma\gamma}(\chi_{c2}) \mathcal{B}(\chi_{c2} \to \gamma J/\psi) 
\mathcal{B}(J/\psi \to e^+e^-+\mu^+\mu^-)=13.2\pm1.4(\mathrm{stat}) \pm 1.1(\mathrm{syst})$ eV,
and obtain $\Gamma_{\gamma\gamma}(\chi_{c2})=559 \pm 57(\mathrm{stat}) \pm 48(\mathrm{syst}) \pm 36(\mathrm{br})$ eV.
This result is in excellent agreement with the result of 
$\gamma\gamma$--fusion measurement by Belle 
and is consistent with that of the $\bar{p}p \to \chi_{c2} \to \gamma\gamma$
measurement, when they are both reevaluated using the recent CLEO result for the 
radiative decay $\chi_{c2} \to \gamma J/\psi$.
\end{abstract}

\maketitle

The P-wave states of charmonium ($^3P_J$,$^1P_1$) have \mbox{provided} valuable 
information about the $q\bar{q}$ interaction and QCD.   
The two-photon decays of the positive C-parity states $(^{3}P_{J})$
are particularly interesting because at lowest order the two-photon decay of charmonium
is a pure QED process akin to the two--photon decay of positronium. Their study
can shed light on higher order relativistic and QCD radiative corrections.  

The measurement of the two--photon width of $\chi_{c2}$, the $^3P_2$ state of charmonium, 
$\Gamma_{\gamma\gamma}(\chi_{c2})\equiv\Gamma(\chi_{c2}\to\gamma\gamma)$, has a very checkered 
history, with large differences in results from measurements using different 
techniques.  The pre-1992 $\gamma\gamma$--fusion  measurements of $\Gamma_{\gamma\gamma}(\chi_{c2})$ were  
inconclusive, with most of them only establishing upper limits of several keV. 
In 1993, the E760 experiment at Fermilab reported the result 
from their $\bar{p}p\to\chi_{c2}\to\gamma\gamma$ measurement, $\Gamma_{\gamma\gamma}(\chi_{c2})=320\pm80\pm50$ eV \cite{e760}, a factor of more than 3 smaller than the smallest 
limit established by the $\gamma\gamma$--fusion measurements. The 
$\gamma\gamma$--fusion experiments continue to report much larger valuers of 
$\Gamma_{\gamma\gamma}(\chi_{c2})$ than the $\bar{p}p$ experiments,  
with the result of the $\gamma\gamma$--fusion measurement from Belle \cite{belle}, $\Gamma_{\gamma\gamma}(\chi_{c2})=850\pm127$ eV, which 
is still three times larger than the latest $\bar{p}p$ measurement of Fermilab E835 
\cite{e835}, $\Gamma_{\gamma\gamma}(\chi_{c2})=270\pm59$ eV.  It is this continuing discrepancy between the present good-statistics 
measurements that has motivated the investigation reported here.

In this investigation we report on a measurement of 
$\Gamma_{\gamma\gamma}(\chi_{c2})$ by the study of
the $\gamma\gamma$--fusion reaction
\begin{equation}
e^+e^-\to e^+e^-(\gamma\gamma) \;,\; \gamma\gamma\to\chi_{c2}\to\gamma J/\psi\to\gamma l^+l^-.
\end{equation}

The data sample used for the analysis was taken at the Cornell Electron Storage Ring
(CESR) with the detector in the CLEO III configuration \cite{cleo3}. The detector
  provides 93$\%$ coverage of solid angle for charged and
neutral particle identification. The detector components important for this analysis
are the drift chamber (DR) and CsI(Tl) crystal calorimeter (CC). The DR and CC are
operated within a 1.5 T magnetic field produced by a superconducting solenoid. 
The DR detects charged particles and measures their
momenta. The CC allows measurements of electromagnetic showers with energy resolution $\sigma(E)/E=2.3-2.7\%$ for $E_\gamma=0.3-0.6$ GeV.

The data consist of a 14.4 fb$^{-1}$ sample of $e^{+}e^{-}$ collisions at or near
the energies of $\Upsilon(1S-5S)$ resonances and around the $\Lambda_{b} 
\overline{\Lambda_{b}}$ threshold in the range of center--of--mass energies $\sqrt {s} = 9.46 - 11.30$ GeV.  The data sample sizes are given in Table I.

The two--photon partial width $\Gamma_{\gamma \gamma}(\chi_{c2})$  was measured in untagged $\gamma\gamma$--fusion reaction of Eq. (1). Events with $\gamma e^{+}e^{-}$ or $\gamma \mu^{+}\mu^{-}$
in the final state were selected. The selected events are required to have two charged tracks and zero net charge. All charged particles were required to lie within the drift chamber volume and 
satisfy standard requirements for track quality and distance of closest approach 
to the interaction point. 

The photon produced in the decay $\chi_{c2} \to \gamma l^{+}l^{-}$ typically
has an energy $E_{\gamma}\approx0.46$ GeV. 
The selected events are required to have only one electromagnetic shower with an 
energy $0.3<E_{\gamma}<0.6$ GeV, and to be isolated from the
nearest charged track by an angle $>20^{\circ}$. The total energy of remaining 
electromagnetic showers in the event $E_{\mathrm{tot}}(\mathrm{neut})$ was required to be $<0.3$ GeV. 

\begin{table}[t!]
\caption{
Data used in the present analysis. Average values of $\sqrt{s}$ and corresponding luminosities are listed.
}
\begin{center}
\begin{tabular}{c|cccccc}
\hline \hline
Data & $\Upsilon(1S)$ & $\Upsilon(2S)$ & $\Upsilon(3S)$ & $\Upsilon(4S)$ & $\Upsilon(5S)$ & $\Lambda_{b} \overline{\Lambda_{b}}$ \\
\hline
$\mathcal{L}$ (fb$^{-1}$) & 1.399 & 1.766 &  1.598 & 8.566 & 0.416 & 0.688 \\
$\sqrt{s}$ (GeV) & 9.458 & 10.018 & 10.356 & 10.566 & 10.868 & 11.296 \\
\hline \hline
\end{tabular}

\end{center}
\end{table}

The total energy of the system, $E_{\mathrm{tot}}(\gamma l^{+}l^{-})$, defined as the energy 
sum of the lepton pair and the candidate photon, was required to be 
less than 5 GeV.
This cut has an efficiency of $\sim 96\%$ and removes all background that arises when $\psi(2S)$ is produced via initial state radiation, decays to
$\gamma\gamma J/\psi$, and one photon is not detected. 

Untagged $\gamma\gamma$--fusion events are characterized with small transverse 
momentum; therefore $p_{\mathrm{tot}}^{\perp}(\gamma l^{+}l^{-})<0.15$ GeV/$c$
was required.

Lepton pairs of low transverse momentum may also be directly produced by
$\gamma\gamma$--photon fusion. These constitute a background that is removed by rejecting
lepton pairs with $p_{\mathrm{tot}}^{\perp}(l^{+}l^{-})<0.1$ GeV/$c$.

To identify two charged tracks as electrons or muons, the $E/p$ variable was used, 
where $E$ is the energy determined from the calorimeter and $p$ is the momentum
determined from track reconstruction. For both muons $0<E/p<0.3$ is required, and
for both electrons  $0.85<E/p<1.15$ is required.  If a photon of energy larger than 0.03 GeV is present within a $5^{\circ}$ angle cone
around the lepton direction, it is assumed to be the result of bremsstrahlung
and its momentum is added to the momentum of the track.

The signal Monte Carlo (MC) sample for untagged $\gamma\gamma$
fusion production of $\chi_{c2}$ resonance was generated
using the $\gamma \gamma$ fusion formalism from Budnev $et ~al.$ \cite{budnev}. 
MC samples were produced for each value of 
$\sqrt{s}$ listed in Table I. For the calculation of the overall
event selection efficiencies, MC samples were weighted according to the luminosity 
of each data set. 

According to Budnev et al. \cite{budnev}, when the photons are 
transversely polarized, the untagged $\gamma \gamma$ production cross section is related to the 
two--photon cross section by
\begin{displaymath}
d\sigma_{e^{+}e^{-} \to e^{+}e^{-}\chi_{c2}} = d{\cal L}_{\gamma\gamma}^{TT}(W^{2})
\sigma_{\gamma\gamma \to \chi_{c2}}^{TT},
\end{displaymath}
where ${\cal L}_{\gamma\gamma}^{TT}$ is the $\gamma\gamma$ luminosity function and $W$ is the two--photon invariant mass.  The implementation of the above formalism has been discussed in detail by Dominick et al. \cite{cleob}, and we follow it.  We calculate $\sigma(\chi_{c2})/\Gamma_{\gamma\gamma}(\chi_{c2})$
for each value of $\sqrt{s}$ assuming that $\chi_{c2}$ production
in the fusion of two transverse photons is significant only be in the helicity 2
state \cite{licb}, and that the radiative transition
$\chi_{c2} \to \gamma J/\psi$ is pure E1. We assume that the intermediate vector 
meson in the Budnev formalism is a $J/\psi$, and we implement the proper angular
distribution \cite{cleob} in calculating efficiencies. The luminosity-weighted
average value of $\sigma(\chi_{c2})/\Gamma_{\gamma\gamma}(\chi_{c2})$ is determined to be 4.93 pb/keV.

Good agreement is observed between the data and MC distributions for
$E_{\mathrm{tot}}(\gamma l^{+}l^{-})$, $p_{\mathrm{tot}}^{\perp}(\gamma l^{+}l^{-})$, $E_{\gamma}$, $l^+,~l^-$ momenta, $E_{\mathrm{tot}}(\mathrm{neut})$, $E/p$ for leptons, and the photon and lepton 
angular distributions.  The $E/p$ and  $p_{\mathrm{tot}}^{\perp}(\gamma l^{+}l^{-})$ distributions are shown in Fig. 1, and the lepton and photon angular distributions are shown in Fig. 2.

\begin{figure}[!b]
\begin{center}
\includegraphics[width=3.3in]{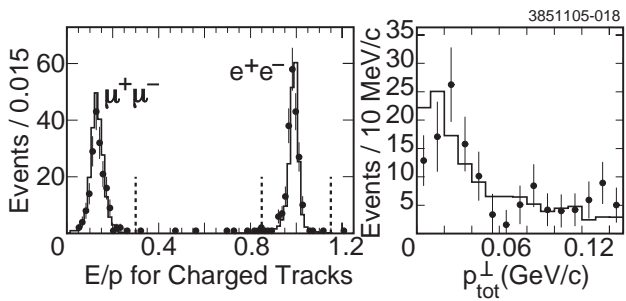}
\caption{
Distributions of $E/p$ (left) and $p_{\mathrm{tot}}^{\perp}$ (right) in data (points) and in the signal MC (histograms). Vertical dashed lines
indicate the $E/p$ cuts for electrons and muons.
}
\end{center}
\end{figure}

\begin{figure}[!b]
\begin{center}
\includegraphics[width=3.2in]{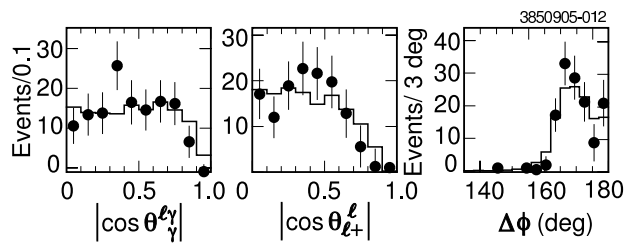}
\caption{
Comparison of the background subtracted data (points) and the signal MC
(histograms)
distributions of $|\cos\Theta_{\gamma}^{l\gamma}|$, $|\cos\Theta^{l}_{l^+}|$ and
$\Delta \phi$. $\Theta_{\gamma}^{l\gamma}$ is the polar angle of the photon in
the $l^{+}l^{-}\gamma$ rest frame, $\Theta^{l}_{l^+}$ is the polar angle of the
positive lepton in the $l^{+}l^{-}$ rest frame, and  $\Delta \phi$ is the azimuthal
angle difference between the momenta of the two leptons in the laboratory frame.
}
\end{center}
\end{figure}

For the data at each $\sqrt{s}$, efficiencies of all event selection requirements were determined from signal MC simulations, and the averages
are listed in Table II.

\begin{table}[!t]
\caption{
Efficiencies of the different event selection criteria.
}
\begin{center}
\begin{tabular}{lcc}
\hline \hline
Selection cut & $e^{+}e^{-}$ & $\mu^{+}\mu^{-}$ \\
& Channel $(\%)$ & Channel $(\%)$ \\
\hline
N(charge)=2 & 68.9 & 70.8 \\
Total Charge=0 & 98.7 & 98.7 \\
Only one $\gamma$ with & & \\
$0.3<E_{\gamma}<0.6$ GeV & 52.8 & 53.7 \\
Lepton $E/p$ & 92.4 & 98.3 \\
$E_{\mathrm{tot}}(\gamma l^{+}l^{-})<5$ GeV & 96.1 & 95.3 \\
$E_{\mathrm{tot}}(\mathrm{neut})<0.3$ GeV & 99.0 & 99.1 \\
$p_{\mathrm{tot}}^{\perp}(l^{+}l^{-})>0.1$ GeV/$c$ & 99.0 & 98.9 \\
$p_{\mathrm{tot}}^{\perp}(\gamma l^{+}l^{-})<0.15$ GeV/$c$ & 62.1 & 62.4 \\
$M(l^{+}l^{-})=M(J/\psi) \pm 30$ MeV & 81.9 & 93.0 \\
Trigger & 97.5 & 85.7 \\ \hline
Overall efficiencies & 15.5 & 17.1 \\ \hline \hline
\end{tabular}
\end{center}
\end{table}

A two dimensional plot of the lepton pair masses $(e^+e^-+\mu^+\mu^-)$ versus the mass difference $\Delta M=M(\gamma l^{+}l^{-})-M(l^{+}l^{-})$ is shown in Fig. 3.  It shows a clear enhancement at the mass of $J/\psi$.  A cut $M(l^+l^-)=M(J/\psi)\pm30$ MeV was therefore used.  The resulting distributions of $\Delta M$ are shown in Fig. 4 for (a) $e^+e^-$, (b) $\mu^+\mu^-$ and (c) $e^+e^-$ plus $\mu^+\mu^-$.

\begin{figure}[!b]
\begin{center}
\includegraphics[width=2.6in]{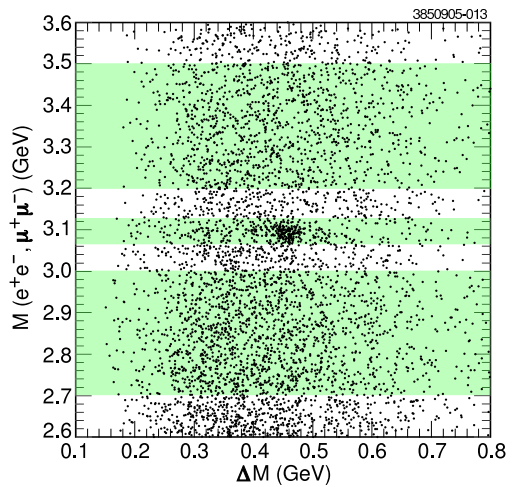}
\caption{
Scatter plot of the $\Delta M = M(\gamma l^{+}l^{-})-M(l^{+}l^{-})$ with respect 
to the two lepton effective mass in data.
Top and bottom horizontal shaded bands are the areas defined as the $J/\psi$ sideband 
regions, and the middle band is the area defined as the $J/\psi$ signal region. 
}
\end{center}
\end{figure}

\begin{figure}[h]
\begin{center}
\includegraphics[width=2.6in]{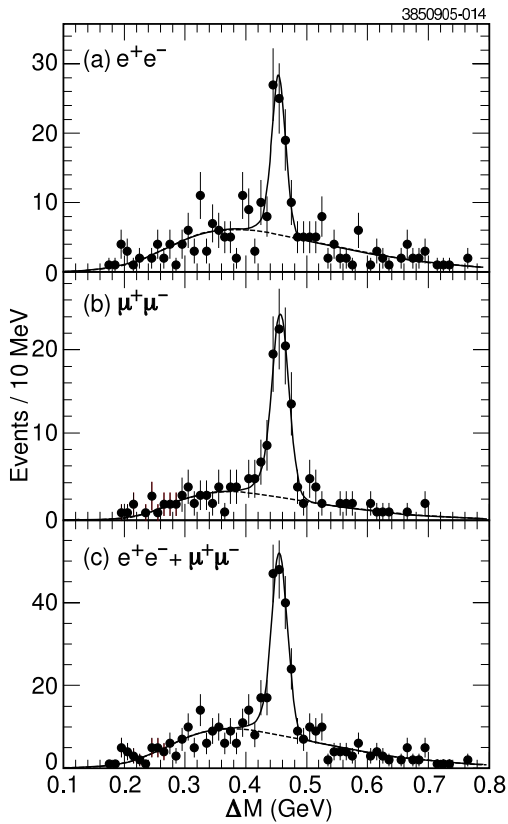}
\caption{
$\Delta M=M(\gamma l^{+}l^{-})-M(l^{+}l^{-})$ mass difference
distributions for $e^{+}e^{-}$ channel (a), 
$\mu^{+}\mu^{-}$ channel (b), and the sum (c).
The solid line curves are results of fit to the data points using the background indicated by the dashed line curves.
}
\end{center}
\end{figure}

Three different methods, all using the background shape determined from the $J/\psi$
sideband region [$M(l^{+}l^{-})$=2.7--3.5 GeV, omitting $M(l^{+}l^{-})$=3.0--3.2 GeV],
were used. Fits using the Crystal Ball line shape (which consists of a Gaussian with a low energy tail) \cite{cbshape}, signal MC peak shape, and simple counts in the region $\Delta M$=0.42--0.49 GeV led to yields and efficiencies that differ by less than $\pm$2$\%$. 

The observed yield, $N_{\mathrm{obs}}$, leads to
\begin{displaymath}
\Gamma_{\gamma\gamma}(\chi_{c2})\times \mathcal{B}(\chi_{c2}\to\gamma l^+l^-) =\frac{N_{\mathrm{obs}}}{\epsilon \mathcal{L} (\sigma(\chi_{c2})/\Gamma_{\gamma\gamma}(\chi_{c2}))},
\end{displaymath}
where $\epsilon$ is the total efficiency, $\mathcal{L}$ is the total luminosity of the data
used, $\sigma(\chi_{c2})/\Gamma_{\gamma\gamma}(\chi_{c2})=4.93$ pb/keV as determined earlier, and
$\mathcal{B}(\chi_{c2} \to\gamma l^{+}l^{-})=\mathcal{B}_{1}(\chi_{c2} \to \gamma J/\psi) \mathcal{B}_{2}(J/\psi \to l^{+}l^{-})$.

In Table III we present our results, which are averages of the results for the three different signal yield extraction methods.  We present the results for $e^+e^-$ and $\mu^+\mu^-$ separately, and for their sum.  Our directly determined result for the sum is
\begin{displaymath}
\Gamma_{\gamma\gamma}(\chi_{c2}) \mathcal{B}(\chi_{c2}\to\gamma(e^+e^- + \mu^+\mu^-))=13.2\pm1.4 ~\mathrm{eV}.
\end{displaymath}
We use $\mathcal{B}_1(\chi_{c2}\to\gamma J/\psi)=(19.9\pm1.3)\%$ as measured by CLEO \cite{cleo}, and $\mathcal{B}_2(J/\psi\to e^+e^-)=\mathcal{B}_2(J/\psi\to \mu^+\mu^-)=5.95\pm0.06)\%$ \cite{cleoo}, to obtain $\Gamma_{\gamma\gamma}(\chi_{c2})$ listed in Table III.  Various sources of systematic uncertainty were studied.  These are listed in Table IV, and are combined in quadrature to give a total systematic error of $\pm8.6\%$.  
Thus, our final result is
\begin{displaymath}
\Gamma_{\gamma\gamma}(\chi_{c2})=559 \pm 57(\mathrm{stat}) \pm 48(\mathrm{syst}) \pm36(\mathrm{br}) ~\mathrm{eV}.
\end{displaymath}

\begin{table}[!t]
\caption{
Average of results for the three signal count extraction methods.
}
\begin{center}
\begin{tabular}{cccc}
\hline \hline
$l^+l^-$ & $N_{\mathrm{obs}}$ & $\Gamma_{\gamma\gamma}(\chi_{c2})\mathcal{B}(\chi_{c2}\to\gamma l^{+}l^{-})$ & $\Gamma_{\gamma\gamma}(\chi_{c2})$ \\
 & & (eV) & (eV) \\ \hline
$e^{+}e^{-}$ & 68$\pm$11 & 6.4$\pm$1.0 & 544$\pm$87 \\
$\mu^{+}\mu^{-}$ & 79$\pm$11 & 6.8$\pm$0.9 & 571$\pm$76 \\
Total & 147$\pm$15 & 13.2$\pm$1.4 & 559$\pm$57 \\ \hline \hline
\end{tabular}
\end{center}
\end{table}

\begin{table}[!t]
\caption{
Sources of systematic uncertainties.
}
\begin{center}
\begin{tabular}{lc}
\hline \hline
Source & Systematic uncertainty ($\%$) \\ \hline
integrated luminosity, $\mathcal{L}$ & $\pm$3.0 \\
trigger efficiency & $\pm$3.0 \\
signal yield extraction & $\pm$1.3 \\
$J/\psi$ line shape modeling & $\pm$1.6 \\
photon resolution modeling & $\pm$1.3 \\
event selection & $\pm$4.8 \\
tracking & $\pm$2.0 \\
photon finding & $\pm$2.0 \\
$J/\psi$ (versus $\rho,\;\phi$) in $\gamma\gamma$ & $\pm$3.0 \\ 
pure E1 (versus E1 + 10\% M2) & $\pm3.0$ \\
\hline
overall & $\pm$8.6 \\ 
\hline \hline
\end{tabular}
\end{center}
\end{table}

Table V shows a compilation of the published results of Belle \cite{belle}, E835 \cite{e835}, and CLEO \cite{cleoc}, together with our result.  We find that a large part of the discrepancy between the earlier $\gamma\gamma$--fusion
results and the $\bar{p}p \to \chi_{c2} \to \gamma\gamma$ results arises from
the use of the old values of $\mathcal{B}(\chi_{c2} \to \gamma J/\psi)$.  This possibility was indeed anticipated by Belle \cite{belle}. Both 
Belle \cite{belle} and E835 \cite{e835} used the 2000 PDG value of $\mathcal{B}(\chi_{c2} \to \gamma J/\psi)=(13.5\pm1.1)\%$. As shown in Table V, when these results are reevaluated using $\mathcal{B}(\chi_{c2} \to 
\gamma J/\psi)=(19.9 \pm 1.3)\%$, as recently measured by CLEO \cite{cleo},
the Belle result \cite{belle} comes into complete agreement with ours, and even the latest $\bar{p}p$ result \cite{e835} 
becomes statistically consistent with ours.

Many theoretical predictions based on potential model calculations exist in the
literature. In an early relativistic calculation, Barnes \cite{barnes} predicted $\Gamma_{\gamma\gamma}(\chi_{c2})=560$ eV.  In calculations including both relativistic and one--loop QCD radiative corrections, Gupta et al. \cite{gupta} and Ebert et al \cite{ebert} predict $\Gamma_{\gamma\gamma}(\chi_{c2})=570$ eV and 500 eV, respectively.  All three predictions are in agreement with our result.

An estimate of the strong coupling constant, $\alpha_S(m_c)$ can be obtained by comparing $\Gamma_{\gamma\gamma}(\chi_{c2})$ with
$\Gamma_{gg}(\chi_{c2})$. With the known first order QCD radiative corrections for the 
two widths \cite{rosner}, the pQCD prediction is
\begin{equation}
{\Gamma_{\gamma\gamma}(\chi_{c2}) \over \Gamma_{gg}(\chi_{c2})} =
{8\alpha^{2} \over 9\alpha_{s}^{2}} \times \left( {1-{5.33 \over \pi}\alpha_{s} \over 1-{2.2 \over \pi}\alpha_{s}} \right).
\end{equation}

\begin{table}[!t]
\caption{
Comparison of our result for $\Gamma_{\gamma\gamma}(\chi_{c2})$ with
the results of the two recent $\gamma\gamma$--fusion measurements and the Fermilab E835
$\bar{p}p$ experiment. The second column gives the results as published and the third
column gives the result after reevaluation using the CLEO
measured values for $\mathcal{B}_{1}(\chi_{c2} \to \gamma J/\psi)$ and $\mathcal{B}_{2}(J/\psi \to
l^{+}l^{-})$ \cite{cleo},\cite{cleoo}. Also, the average Fermilab measured value of
$\Gamma_{\mathrm{tot}}(\chi_{c2})$ \cite{e835wid} is used to recalculate the E835 result \cite{e835},
and PDG2004 value of $\mathcal{B}(\chi_{c2} \to 4\pi)$ \cite{pdg} is used to recalculate the CLEO result \cite{cleoc}.
}
\begin{center}
\renewcommand{\arraystretch}{1.2}
\begin{tabular}{ccc}
\hline \hline
Experiment [Ref.] & $\Gamma_{\gamma\gamma}(\chi_{c2})$ (eV) & $\Gamma_{\gamma\gamma}(
\chi_{c2})$ (eV) \\
Quantity Measured & (as published) & (as reevaluated) \\ \hline
Present: $\gamma\gamma\to\chi_{c2}$ & & \\
$\Gamma_{\gamma\gamma}(\chi_{c2}) \mathcal{B}(\chi_{c2}\! \to\! \gamma l^{+}l^{-})$ & \multicolumn{2}{c}{\bf{559(57)(48)(36)}} \\
\hline
Belle \cite{belle}:  $\gamma\gamma\to\chi_{c2}$ & & \\
$\Gamma_{\gamma\gamma}(\chi_{c2}) \mathcal{B}(\chi_{c2}\! \to\! \gamma l^{+}l^{-})$ & 850(80)(70)(70) &
570(55)(46)(37) \\ \hline
CLEO \cite{cleoc}: $\gamma\gamma\to\chi_{c2}$ & & \\
$\Gamma_{\gamma\gamma}(\chi_{c2}) \mathcal{B}(\chi_{c2}\! \to\! 4\pi)$ & 530(150)(60)(220) & 432(122)(54)(61) \\
\hline
E835 \cite{e835}: $\bar{p}p\to\chi_{c2}$ & & \\
$(\chi_{c2}\!\to\!\gamma\gamma)/(\chi_{c2}\!\to\!\gamma J/\psi)$ & 270(49)(33) & 384(69)(47) \\
 \hline \hline
\end{tabular}
\end{center}
\end{table}

 The hadronic width,
$\Gamma_{gg}(\chi_{c2})=\Gamma_{\mathrm{tot}}(\chi_{c2}) \times \mathcal{B}(\chi_{c2} \to gg)=
\Gamma_{\mathrm{tot}}(\chi_{c2}) \times [1-\mathcal{B}(\chi_{c2} \to \gamma J/\psi)]=1.55 \pm 0.11$ MeV,
obtained by using $\Gamma_{\mathrm{tot}}(\chi_{c2})=1.94 \pm 0.13$ MeV \cite{e835wid} and 
$\mathcal{B}(\chi_{c2} \to \gamma J/\psi)=0.199 \pm 0.013$ \cite{cleo}. Using our result for $\Gamma_{\gamma\gamma}(\chi_{c2})$, we obtain 
$\Gamma_{\gamma\gamma}(\chi_{c2})/\Gamma_{gg}(\chi_{c2})=(361 \pm 59) 
\times 10^{-6}$. Equating this to the pQCD expression (Eq. 2) but not including the QCD radiative 
corrections in the large parentheses, gives $\alpha_{s}(m_{c})=0.36 \pm 0.03$. The pQCD expression with the QCD radiative corrections (Eq. 2) leads to the value $\alpha_{s}(m_{c})=0.29 \pm 0.02$.

We gratefully acknowledge the effort of the CESR staff in providing us with excellent 
luminosity and running conditions. This work was supported by the National Science
Foundation and the U.S. Department of Energy.

\end{document}